\documentclass[11pt,preprint]{aastex}

\newcommand\msun{\rm M_{\odot}}

\newcommand\be{\begin{equation}}
\newcommand\en{\end{equation}}

\newcommand\kms{\rm{\, km \, s^{-1}}}

\newcommand\etal{{\rm et al}.\ }

\begin{document}

\title{Flows, Fragmentation, and Star Formation. 
I. Low-mass Stars in Taurus}

\author{Lee Hartmann}
\begin{abstract}
The remarkably filamentary spatial distribution of young stars in
the Taurus molecular cloud has significant implications for understanding
low-mass star formation in relatively quiescent conditions.  The
large scale and regular spacing of the filaments suggests 
that small-scale turbulence is of limited importance, 
which could be consistent with driving on large scales
by flows which produced the cloud.  The small spatial dispersion
of stars from gaseous filaments indicates that the low-mass stars
are generally born with small velocity dispersions relative to 
their natal gas, of order the sound speed or less.  The spatial
distribution of the stars exhibits a mean separation
of about 0.25 pc, comparable to the estimated Jeans length in the
densest gaseous filaments, and is consistent
with roughly uniform density along the filaments. 
The efficiency of star formation in filaments is much higher
than elsewhere, with an associated higher frequency of
protostars and accreting T Tauri stars.
The protostellar cores generally are
aligned with the filaments, suggesting
that they are produced by gravitational fragmentation,
resulting in initially quasi-prolate cores.  Given the absence
of massive stars which could strongly dominate cloud dynamics,
Taurus provides important tests of
theories of dispersed low-mass star formation and 
numerical simulations of molecular cloud structure and evolution. 
\end{abstract}
\affil{Harvard-Smithsonian Center for Astrophysics, 60 Garden
St., Cambridge, MA 02138;\\
Electronic mail: hartmann@cfa.harvard.edu}


\section{Introduction}

The Taurus-Auriga molecular cloud long has been a touchstone
for studies of star formation.  Although Taurus is not typical
of most (massive) star-forming regions, its proximity and low
extinction mean that its stellar population is the best determined
of any star-forming cloud.  Perhaps more importantly, Taurus
is relatively quiescent, with low turbulent velocities and a lack
of massive stars to dissociate, ionize, and otherwise disrupt
the cloud.  If the simple, static models of the original
paradigm of low-mass star formation (e.g., Shu, Adams, \& Lizano 1987) can 
be applied anywhere, they should work in Taurus.

The spatial distributions with which stars are formed can
provide important clues to the processes of star formation.
Even though Taurus does not contain populous
clusters, many of its stars fall into loose groups (Jones \& Herbig 1979;
Gomez \etal 1993).  So far, the most detailed studies of the stellar spatial
distribution in Taurus have considered the two-point correlation function
(Gomez \etal 1993) and the related mean surface density
of companions (MSDC; Larson 1995).  Larson (1995) found that 
the MSDC exhibited roughly two distinct power-law distributions
at small and large scales, the inner region corresponding to close
binaries and multiple systems, and the outer region representing
the clustering properties of the stars.  A break between these
two distributions was identified at about $\sim 0.04$~pc, which Larson 
suggested was the Jeans length in Taurus
(see also Simon 1997 and Bate, Clarke, \& McCaughrean 1998).

The molecular gas in Taurus has long been recognized to be
filamentary in nature (e.g., Schneider \& Elmegreen 1979;
Scalo 1990; Mizuno \etal 1995).  It is also well-recognized
that the young stars are strongly correlated with gas and dust;
i.e., that the stellar distribution must exhibit filamentary
structure as well.  However, the extent of the stellar 
filamentary distribution, and its physical significance, 
has not been given sufficient attention, nor has it been viewed
in the context of recent numerical simulations of molecular clouds.
The spatial distribution of young objects in Taurus is
reexamined from this point of view, with emphasis on its
relationship to our previous suggestion that the Taurus cloud 
(like other nearby molecular clouds) was formed by large-scale 
flows in the interstellar medium (Ballesteros-Paredes,
Hartmann, \& Vazquez-Semadeni 1999, BHV; Hartmann, 
Ballesteros-Paredes, \& Bergin 2001, HBB).

\section{Spatial distribution}

\subsection{Sample}

The basic sample of young stars in Taurus used in this paper
is taken from the tabulation of Kenyon \& Hartmann (1995) 
with various additions.  A few new low-mass stars were added from
the studies of Brice\~no \etal (1998, 1999).  I further added
a few stars from the Rosat All-Sky Survey (RASS) studies of Wichmann
\etal (1996, 2000).

Some investigators have argued that the
RASS identified large numbers of previously unknown, much older Taurus
members (e.g., Neuh\"auser \etal 1995; Wichmann \etal 1996), a claim
that does not hold up under detailed scrutiny (Brice\~no \etal 1997;
Martin \& Magazzu 1998; HBB).
However, for present purposes any possible older population is unimportant,
because these stars can and probably have dispersed significantly from their 
birthsites, and so their positions reveal little about star formation processes. 
For this reason
neither the ``post-T Tauri'' stars identified on the
basis of Li by Martin \& Magazzu (1998) nor the ``zero age main sequence'' 
stars of Wichmann \etal (2000) have been included.  I have also 
eliminated a number of stars
originally in Kenyon \& Hartmann (1995) which were taken from the study
of Walter \etal (1988), based on low Li abundances and also positions
in the HR diagram indicating ages greatly in excess of 10 Myr.
Note that if all potential pre-main sequence X-ray members
were added to the sample, there would be little effect on the subsequent conclusions, 
because these stars are widely dispersed spatially, and thus constitute 
a small fraction of the objects in the area under consideration.

\subsection{Group structure}

Figure 1 shows the spatial distribution of young stellar objects
superimposed upon the $^{12}$CO map of Megeath, Dame,
\& Thaddeus (2002).  Apart from a few small, localized, widely
distributed groups (the largest of which is the L1551 group),
most of the Taurus stars are found in three nearly
parallel, elongated bands.  Even the
two groups or mini-clusters of L1495 lie at the western end
of the main band.  The stellar bands lie along parallel filaments
in the molecular gas, recognized earlier from extinction studies
(Schneider \& Elmegreen 1979).  (In the following I 
use the term ``band'' to refer to extended
distributions of stars, and use ``filament'' to refer to the
narrower dense gas structures and stars near these densest
structures.)

Figure 2a shows the spatial distributions of the young stars
in the central Taurus region, along with the positions of 
optically-determined cloud cores from the unbiased
survey of Lee \& Myers (1999).  The overall regularity of the
structure is perhaps clearer than in the $^{12}$CO gas.
Three main bands of stars appear to be roughly parallel and 
approximately spaced apart by a projected distance of about 
1.5 degrees, $\sim 3.7$~pc at the 140 pc distance of Taurus
(Kenyon, Dobryzcka, \& Hartmann 1994).  The two groups of
the L1495 region are separated from the main stellar band
by a similar distance.

Figure 2b shows the distributions of objects sorted into groups
somewhat arbitrarily for further analysis; for example,
groups 1 and 2 might be merged into one group.  Comparison
with Figure 2a shows that only a few, widely-dispersed objects 
have been deleted from the sample.  Stars
near the two dense groupings in L1495 have also been marked.
The members of the groups shown in Figure 2b are listed in
Table 1.

Figure 3 shows schematically the approximate positions 
of the $^{13}$CO filaments identified by Mizuno \etal (1995)
superimposed upon the stellar distribution.
The positions of the main groups are correlated with
the main clusters of gaseous filaments.  There is clearly
significant substructure within the main bands, especially
in the regions of Group 1 and L1495, while Groups 3 and 4 show
significantly simpler and much more linear filamentary structure,
consistent with the spatial distribution of the stars, as discussed
further in \S 2.3.

The $^{13}$CO data demonstrate a velocity difference between Groups 1 and 2 
(connecting the clouds B213, L1521, and Heiles Cloud 2), but this
difference may simply reflect the large-scale velocity gradient in Taurus.
For the same reason the filament extending westward from the Group 3
may be part of the same structure.
Several filaments connect the two concentrations of
the L1495 cloud at the western end of the complex, suggesting that
these concentrations should be considered together.

Figure 4 shows the distribution of Class I and optical cores.  
These objects are more closely correlated in space
than the Class II and Class III sources.
In particular, the Group 3 objects centered near (l,b = 174, -16) 
display a remarkably linear structure (associated with the B18 cloud).  
This structure is considered in more detail in the following section.
A section of Group 1 near (l,b = 171, -16) is associated with 
the narrow B213 filament, and also displays a remarkably narrow structure.

\subsection{Ages and spatial dispersion}

It would not be surprising if the members of these spatially-concentrated
groups are in general younger than the rest of the Taurus population.
One simple, likely indicator of age is the spectral energy distribution
class.  One would expect that the Class 0 - I objects (protostars,
with remnant infalling envelopes) are in general younger than the
(mostly) optically-visible T Tauri stars.  In turn, since accretion
disks dissipate over time, one would expect the Class III objects
(stars without accreting disks) to be older than the Class II objects
(T Tauri stars with disk accretion onto the central star).  However,
there are many Class III objects in Taurus with ages similar to those
of Class II stars (Kenyon \& Hartmann 1995), showing that the
Class II/Class III dichotomy is not merely a function of age. 
Here the classes listed in Kenyon \& Hartmann (1995) have been used when
available; otherwise, the 10 \AA\ criterion for the
H$\alpha$ equivalent width has been used to distinguish Classical T Tauri stars (II)
from Weak T Tauri stars (III).

Of the 204 objects in the total Taurus sample, the numbers of Class I/II/III 
objects are 24/108/72.  The same numbers for groups 1-4 are 19/51/18.\footnote{
The assignment of Class I stars was taken from Kenyon \& Hartmann (1995). The
compilation of Chen \etal (1995) lists 26 total Class 0 and I Taurus stars, not
counting the class 0 source IRAM 04191+1522 found by Andr\'e, Motte, \& Bacmann
(1999).  The assignment of Class I stars in the groups agrees with that of
Chen \etal (1995) except for L1527 $=$ 04368+2557, which Chen \etal identify
as a Class 0 source.  These small discrepancies in classification are not
important for the overall discussion.} 
(Here the two class I/II objects in the samples,
HK Tau and Haro 6-13, have been divided into one net Class I
and one net Class II for counting purposes.) 
L1495 by itself has only one class I object; the ratios are 1/26/13.
The total numbers divided by class for the four groups plus L1495 are 20/77/31;
therefore, the numbers are 4/31/41 for the Taurus objects outside of the groups
and L1495.  There is an unsurprising excess of Class I objects in the groups,
since the groups are strongly correlated with the
dense gas where protostars form.
The results also indicate a significant excess of Class III sources 
in the more distributed population, which would suggest a significant 
dispersal of accretion disks over a timescale comparable 
to the median age of Taurus, $\sim 2$~Myr (Palla \& Stahler 2000).

A more direct test of age differences is to plot an HR diagram
for the T Tauri stars.  Figure 5
shows the results for objects for which spectral types and stellar
luminosities are available from either Kenyon \& Hartmann (1995) or
Brice\~no \etal (1998).  There is some suggestion that the non-grouped
stars are older, although much of the effect may be due to inclusion
of the L1551 group, which previously has been found to exhibit a wide
age spread (Gomez \etal 1992).  To examine the possible difference
in ages in the least model-dependent way, I consider the luminosity
distributions of group and non-group stars over a modest range
of effective temperature; the luminosity distribution then corresponds
to an age distribution (e.g., Hartmann 2001).
A two-sided Kolmogorov-Smirnov test of the luminosity distributions of
K7-M1 stars indicates no strongly significant difference between the group
and non-grouped stars, and thus does not support a significant difference
in age for this mass range.  
Any possible age difference between the group and non-group
T Tauri stars needs to be investigated further, with more complete
estimates of stellar luminosities and effective temperatures, along
with additional examination of possible systematic errors (Hartmann 2001).

The presence of such well-defined bands and filaments in the stellar spatial
distribution implies that the stars in the groups must be relatively
young and that the internal velocity dispersions must not be very large.
One can take advantage of the remarkably linear structure of Group 3
to investigate this in a simple way.
A least-squares linear fit to the positions of the Class I
objects and cores in galactic coordinates, first removing 
the one core that deviates strongly
(B18-2; Lee \& Myers 1999), yields the dotted line shown in Figure 4.
The rms dispersion in distance of these 19 objects (13 cores
and 6 Class I or I/II objects) from the line
is 0.0667 degree, or $\sim 0.16$~pc.  This is remarkably small
considering the total length of the distribution is $\sim 3.93$~degree
$\sim 9.6$~pc.  The spatial dispersion
of the 17 Class II and Class III objects from the same line is somewhat 
larger, $\sim 0.148$~degree~$\sim 0.37$~pc; the dispersion of
Class II objects alone is only marginally smaller, $\sim 0.31$~pc.
It should be noted that this difference in spatial dispersion 
between the Class I/cores and the II/III objects is 
marginally significant, only about 2.3 standard deviations. 

Assuming that the difference in spatial dispersion between
the Class I objects and cores on the one hand, and the T Tauri
stars on the other, is real and is due entirely to dispersal and not to
a difference in birth sites, then an upper limit to the velocity
dispersion can be estimated.  Assuming the average age of the Taurus stars 
of $\sim 2$~Myr (Palla \& Stahler 2000; see Hartmann 2001 for the difficulty
in assigning individual ages to young stars), this would imply
that the Class II/Class III objects of group 3 have a velocity
dispersion perpendicular to the filament (in the plane of the sky)
$\lesssim 0.2 \kms$, comparable
to the sound speed at 10 K.  If these stars are systematically younger
than average (see above), the upper limit to the dispersion might be
somewhat larger; on the other hand, the true birthsites of these
objects could be more widely spread than the Class I/core objects.
In any event, the implied small velocity dispersion
emphasizes the quiescent nature of this filament.

\subsection{Surface density analysis}

As mentioned in the Introduction,
the spatial structure of the Taurus YSOs has previously been 
discussed in terms of companions per unit area, either with
the MSDC (Larson 1995; Simon 1997) or the two-point correlation function
(Gomez \etal 1993).   On large scales, Larson (1995) estimated
that the MSDC was approximately a power-law, 
$\propto \theta^{-0.62}$, and suggested that this indicated
hierarchical clustering, perhaps of a fractal nature.
The slope of the MSDC changed dramatically at small scales 
$\lesssim 0.02$~degree $\sim 0.05$~pc, at which point binary 
and multiple systems dominate the rapid rise in the MSDC.
Larson interpreted this break as setting the Jeans length in Taurus.

Figure 6 shows an updated version of the MSDC, including
additional members of Taurus not known earlier.  
As discussed in Simon (1997), it is difficult to estimate errors
for the MSDC or correlation function because the data bins are 
not independent.  The errors in Figure 6 have been crudely estimated 
by dividing the Taurus stars into four samples and calculating the
resulting dispersion (averaged slightly among bins).
Here I concentrate on the structure of the MSDC outside
of the ``binary'' or companion star regime, $\log \theta \geq -2$;
the innermost region has been treated in detail by Simon (1997).

Using smaller spatial bins spanning a larger range of separations 
than considered by Larson (1995) or Simon (1997), the MSDC exhibits
an apparently significant break near $\log \theta = -1$, which is only 
hinted at in the results of Simon (1997) but appears 
more clearly in the results of Gomez \etal (1993) (see their Figure 2).
The large-scale MSDC slope is steeper, $\sim -0.98 \pm 0.04$, 
than found by Larson or Simon.  The difference appears
to be that the previous investigators binned over larger size scales,
and limited the fit to smaller scales, so that their fits 
spanned the turnover at $\log \theta = -1$.

The MSDC, $\Sigma (\theta) \propto \theta^s$, 
is the number of companions per unit area, which
depends not only upon the tendency of stars to group together,
but also the overall geometry of the star-forming region.  
Consider a simple model in which stars are distributed
roughly uniformly along an (infinitely long) filament.
On scales larger than the filament width, the number of companions 
to any given star will increase linearly with increasing
distance, while the (two-dimensional) area increases as 
the square of the distance; thus, the slope of the MSDC
should approach $s \sim -1$.  
On scales smaller than the filament width, if the stars are uniformly
distributed in space, the MSDC will be flat, $s \sim 0$.
If the stars within the filament tend to be more clustered, $s$ will
be negative, while if there is a tendency for the stars to be no closer 
than a certain distance, $s$ can be positive.

I thus suggest that the filamentary structure in Taurus
is mostly responsible for the slope $s \sim -1$ at large separations,
and that the stellar distribution does not exhibit a fractal structure
(see Blitz \& Williams 1997 for a similar conclusion concerning fractal
structure in the gas).
There may be problems with this interpretation on the largest scales, 
because some of the filaments are only a few degrees long; on the other hand,
eliminating all the bins with $\theta >$~1~degree would still yield
a similar slope at the largest scales considered.
The break between the large-scale structure and 
the binary distribution at approximately 0.25 pc probably
indicates the typical filament width, or some average spacing
between stars, or both (\S 3.3).

\section{Discussion}

\subsection{Filaments and turbulence}

One of the major problems in molecular cloud formation and evolution
is the nature and source(s) of turbulent motions in clouds.
Turbulence clearly plays a role in the dynamics of molecular clouds
(e.g., Larson 1981);
however, recent numerical simulations show that MHD turbulent motions decay 
rapidly in molecular clouds, generally on a crossing time 
or free-fall time (Stone, Ostriker, \& Gammie 1998; Mac Low \etal 1998;  
Padoan \& Nordlund 1999; Mac Low 1999).  If clouds last for a ``long
time'', this rapid damping of MHD turbulence requires the continual
injection of turbulent energy to ``support'' the cloud against its
self-gravity.  On the other hand, if clouds do not have long lifetimes
(BHV; HBB; Elmegreen 2000), then turbulence need not be
regenerated (BHV; Elmegreen 2000; HBB; Pringle, Allen, \& Lubow 2001).
Locally-generated, stellar-driven turbulence
must be important in many regions, especially those in which
high-mass stars are formed.  Taurus would seem to be a good candidate
for minimal effects of local turbulence because it lacks high-mass stars, 
though low-mass stars may inject substantial amounts of energy 
through their bipolar outflows (Matzner \& McKee 1999, 2000).

Although the general filamentary structure of gas and stars
in Taurus has been recognized for some time, its extent
is remarkable; the main band (groups 1 and 2) 
extends essentially from one end of the complex to the other.  
Furthermore, the other bands are roughly parallel to the main
structure.
This large-scale organization of the Taurus clouds strongly
suggests that any turbulence present must be dominated by
a large-scale component, or alternatively, not driven mostly
on small scales, a conclusion reinforced by
recent numerical simulations of molecular clouds.  Many 
simulations often adopt an {\em ad hoc} driving force 
or velocity field, or an initial spectrum of
density fluctuations, with a fixed distribution of amplitudes 
and spatial wavenumbers (e.g., Heitsch, MacLow, \& Klessen 2001;
Ostriker, Stone, \& Gammie 2001; Klessen \& Burkert 2000).  
Consider specifically the simulations of Klessen \& Burkert (2000, 2001) 
and Klessen (2001), who studied the dynamical evolution
of an isothermal molecular cloud in a cubical volume,
with various assumptions about initial density fluctuations
and driving of turbulence on a variety of scales.  
Unsurprisingly, in the Klessen \& Burkert (2000, 2001) simulations
with density fluctuations leading to velocity perturbations,
or in the Klessen (2001) simulations with energy driving,
calculations with significant power in small spatial wavenumbers 
(on small spatial scales) lead to small-scale
structure, while simulations with driving at low wavenumbers (large spatial
scales) lead to larger-scale structure, in particular
the formation of filaments with large 
spatial extent relative to the size of the computational volume
(for example, Figure 1 of Klessen 2001).  

Based on these simulations, as well as simple considerations,
it seems unlikely that local turbulent energy injection has produced the
bands and filaments of Taurus.  Rather, I conjecture that the Taurus 
cloud structure is consistent with large-scale driving of turbulence. 
If, as BHV and HBB suggested, large-scale flows 
are responsible for the formation of the Taurus molecular cloud, 
such large-scale driving could produce large-scale turbulence
and complex-wide structure.  These conclusions are in agreement with the
suggestion of Burkert \& Mac Low (2002) 
that such large-scale driving is dominant in many molecular clouds.
This is also consistent with observations of low-density, 
non-self-gravitating molecular clouds, in which structure is
observed (sometimes filamentary) that must be driven by
external flows (Falgarone \etal 1998; LaRosa, Shore, \& Magnani 1999;
Sakamoto 2002).

The large-scale magnetic field in the Taurus region is well-ordered,
and oriented roughly perpendicular (in projection) to the filaments
(Moneti \etal 1984; Heyer \etal 1987).  This is also consistent
with large-scale, ordered turbulence, as simulations with comparable
magnetic and flow energies show that the turbulent motions
carry the field around as much as the field channels the flow
(see discussion in HBB).  In this way
the magnetic field may provide a preferential direction for concentration. 

In general, one might expect that colliding flows tend to produce
more sheet-like distributions than filaments, although both
are possible results of turbulence (e.g., Klein \& Woods 1998).  
One way of producing filaments from sheets, suggested by
Schneider \& Elmegreen (1979), and pointed out to the author by 
P. Myers (personal communication), is by gravitational fragmentation
(e.g., Miyama \etal 1987a,b; Nakajima \& Hanawa 1996).  
Using the extreme limiting case of
an isothermal, static, self-gravitating sheet, the critical wavelength is 
$\sim 2 \pi H$, where $H$ is the scale height (Larson 1985).
The $\sim 3.7$~pc separation of the filaments would then suggest
a full thickness of the initial sheet of only $\sim 1$~pc.
The sheet thickness could be even less if one assumes that the filaments
are produced by the fastest growing modes, not the critical
one (e.g., Miyama \etal 19897a,b; Nakajima \& Hanawa 1996).
As the Taurus molecular gas is not static but contains supersonic
turbulence, and as it is highly unlikely to be in hydrostatic equilibrium,
these estimates is highly uncertain.
Nevertheless, if the thickness of the sheet or individual filaments
were much greater than 1 pc, they would have to be oriented almost precisely
along the line of sight.  This seems unlikely, particularly for the
Group 3 structure; an intrinsically filamentary structure seems more
probable (e.g., Schneider \& Elmegreen 1979).

In addition, if the clouds are produced by oblique flows, the resulting
turbulent velocity dispersion in the post-shock flow might well be
anisotropic, providing a preferential direction for gravitational
contraction.  Numerical simulations are needed to test this possibility
for enhancing formation of filaments.

\subsection{Filaments and fragmentation}

Broadly speaking, there are two extreme views of how protostellar
cores are fragmented from their surrounding molecular cloud.
One posits gravitational fragmentation into Jeans-mass-sized
objects (e.g., Larson 1985); the other supposes that turbulent
flows concentrate mass into cores which then are gravitationally-bound
(e.g., Padoan \& Nordlund 1999), or that the cloud has
a highly complex, hierarchical if not fractal structure (Elmegreen 1997)
not primarily driven by gravity.
The true situation must fall somewhere in between these extremes,
as star-forming clouds exhibit supersonic turbulence which must
be damped to some extent before gravitational collapse can
proceed, and any concentrations of gas formed by turbulence
will only form stars if they satisfy a Jeans criterion.

Larson (1985) emphasized that gravitational fragmentation generally 
requires the presence of substructure, such as sheets or filaments,
to provide a smaller scale length than the entire cloud.
The fragmentation length is generally a small multiple of a 
characteristic scale length in the gas (e.g., Oganesyan 1960a,b;
Chandrasekhar \& Fermi 1953).  In an equilibrium
self-gravitating sheet or filament, for example, the scale height
is by definition the length scale over which pressure forces
can resist gravity; the Jeans length along the sheet or filament
then must be a few times larger than the scale height.

The filamentary structure of the stellar distribution in
Taurus suggests that gravitational fragmentation in the
filaments should be considered as a mechanism for forming
cores.  The extreme contrary view, in which flows create isolated
cores, is difficult to fit in with the observed linear morphology 
of Taurus; one would expect much more distributed,
``frothy'' or bubble-shaped structures.  It should be acknowledged
that some of the stars in Taurus may result from such flows,
particularly in the more distributed areas, and perhaps such
flows are responsible for formation of tighter groups
such as the L1551 region; however, even some of the distributed
regions do appear to be elongated, and there is evidence in
the $^{12}$CO map for an extended gaseous filament to the west
of L1551 (see Figure 1).

Although the Taurus filaments exhibit supersonic motions, it is
useful to begin by reviewing the properties of isothermal, 
infinite, self-gravitating cylinders as a starting point.
Consider a cylinder infinitely
extended in the $z$ direction, with $R$ the radial distance in
cylindrical coordinates.  The critical line density (mass per unit
length in the z direction) is a function only of temperature (Ostriker 1964); 
\be
m ~=~ 2 c_s^2/G\,,
\en
where $c_s$ is the isothermal sound speed.  Assuming a mean
molecular weight of $2.36 m_H$, this line density corresponds
to
\be
m ~=~ 16.3 \, T_{10}\, \msun \,{\rm pc}^{-1}\,, \label{eq:linecrit}
\en
where $T_{10}$ is the gas temperature in units of 10 K.
Line densities above this critical value result in collapse in the
$R$ direction.  The Jeans length in the $z$ dimension for
the cylinder of critical line density is (Larson 1985) 
\be
\lambda_c ~=~ 3.94 \, c_s^2 / (G \Sigma_0)
\en
or
\be
\lambda_c ~=~ 1.5 \,  T_{10}\,  A_V^{-1}\,{\rm pc} \,. \label{eq:lambdac}
\en
Here $A_V$ is the visual extinction through
the center of the filament, and $\Sigma$ is the corresponding surface density,
where I use the conversion
$A_V = \Sigma/ 4.4 \times 10^{-3}\, {\rm g \, cm^{-2}}$.
The corresponding critical mass for this fragment is (Larson 1985)
\be
M_c ~=~ 24 \, T_{10}^2 \, A_V^{-1}\, {\msun} \,. \label{eq:mcrit}
\en

The filament has a density structure as a function of cylindrical radius R
of
\be
\rho ~=~ \rho_0 \, ( 1 ~+~ R^2/(4 H^2))^{-2}\,,
\en
where the scale height $H$ is given by
\be
H ~=~ c_s^2/(2 G \Sigma_0) ~=~ 0.19 \, T_{10} \, A_V^{-1}\, {\rm pc} \,. \label{eq:h}
\en
A fragmenting filament will clearly be elongated in the
$z$ direction, at least initially (Oganesyan 1960b; Chandrasekhar \& Fermi
1953; Larson 1985).
In terms of the diameter of the filament at which the density
falls to one-half its central value, $d(\rho) = 2.57 H$,
the aspect ratio is $\lambda_c/d(\rho) = 3.09$.  In terms
of the half-mass diameter $d(m) = 4 H$ (e.g., Ostriker 1964), the
aspect ratio is $\lambda_c/d(m) \sim 2$.  Note that the aspect ratio
is independent of temperature and density.

It is also of interest to consider the linear growth time for gravitational
instability.  Larson (1985) extrapolated the results from the incompressible
filament to estimate a rough linear growth rate for the isothermal
cylinder.  From Larson's Figure 1 the growth timescale is estimated to be
\be
\tau ~\sim~ 3.7 \, T_{10}^{1/2} \, A_V^{-1}\, {\rm Myr}\,. \label{eq:tgrowth}
\en

Now compare this limiting static model with the real Taurus filaments and bands.
Table 2 lists lengths and number densities of stars
in the several groups.  For the purpose of determining line number
densities for the main regions of the groups, the westernmost member
of group 3 and the easternmost one and two members of groups 2 and 4,
respectively, have been eliminated.  The resulting line densities
are remarkably similar, $\sim 4$~stars~pc$^{-1}$; only the L1495 group(s)
exhibit much higher density.  This line density implies an average distance
to the nearest (stellar) neighbor of about 0.25 pc; this is similar to
the estimated $\sim 0.3$~pc for the median nearest-neighbor distance 
for all of Taurus found by Gomez \etal (1993), and is consistent
with the break seen in the MSDC in Figure 6.

Masses are not available for all of the individual objects 
(for example, the Class I sources), and in any event
mass determinations are subject to uncertainties in evolutionary tracks
and binary/multiple companion masses.
For the purpose of a rough comparison, it suffices to estimate the
average system mass as about $1 \msun$, a number arrived at by adopting
the typical Taurus stellar mass of $\sim 0.7 \msun$ and recognizing that
most objects in Taurus are members of multiple systems (Simon \etal 1995).
The stellar bands in Taurus thus have a typical line
density of roughly $4 \, \msun \, {\rm pc}^{-1}$, approximately 1/4 of the
the critical line density of the equilibrium isothermal filament
at 10 K.  

The stars in the groups constitute less mass than the
associated gas.  To estimate this, I turn to the C$^{18}$O survey
of Taurus by Onishi \etal (1996, 1998).  Adding up the masses of
C$^{18}$O cores identified in the B18 cloud (corresponding to the
main region of Group 3) results in a total of $126 \msun$ spanning
about 3.6 pc, or about $35 \, \msun \, {\rm pc}^{-1}$.  As a check,
a similar density $\sim 30 \, \msun \, {\rm pc}^{-1}$ is found for the B213
filament (in Group 1).  These values are approximately twice the
critical density for the equilibrium isothermal filament at 10 K.
(Note that Onishi \etal [1996] estimate a possible uncertainty of a factor
of two in conversion from the C$^{18}$O measurements to total
molecular hydrogen density.)

The difference between the average line densities of the stellar
groups, which (outside of L1495) are about 1/4 of the critical value 
for a static isothermal cylinder, and that of the dense
gas at about twice the critical value, indicates the limited efficiency
of star formation at the present epoch.   As star formation is clearly still
continuing, given the large proportion of Class I sources in the groups,
the stellar density may increase from the present value.  If one adopts
the estimate that the average age of the Taurus stars is $\sim 2$~Myr (Palla \&
Stahler 2000), and that
molecular clouds in the solar neighborhood disperse after a timescale
of order 4~Myr or less (HBB), it is possible that Taurus is only about
halfway through forming stars; in this case the ultimate stellar
density in the groups might approach half the critical value, an
efficiency of conversion of $^{18}$CO dense filamentary gas to stars
of order 25\%. 

The average column density of the starless cores,
as estimated by Onishi \etal (1998), is $\sim 5 \times 10^{21} {\rm cm}^{-2}$, or
$A_V \sim 5$.  Inserting this value into equations
(\ref{eq:lambdac}), (\ref{eq:mcrit}), and (\ref{eq:tgrowth})
yields $\lambda_c \sim 0.3 \, {\rm pc}$,
$M_c \sim 5 \, \msun$,
and $\tau \sim 0.7$~Myr, all assuming hydrostatic equilibrium at 10 K.
These size scales, masses, and lifetimes are comparable to
the average properties estimated for NH$_3$ or optical cores in Taurus 
(Jijina, Myers, \& Adams 1999; Lee \& Myers 1999).

This picture is consistent with the structure of the MSDC.  Although Larson (1995) interpreted
the break at $\sim 0.02$~degree $\sim 0.05$~pc as the Jeans length in Taurus, it seems
more reasonable to interpret the break in the MSDC at 0.25 pc scale as
the true average Jeans length.  This is roughly consistent with the results
of Blitz \& Williams (1997), who found a characteristic length
scale in $^{13}$CO emission in Taurus of order 0.25-0.5 pc, and similarly
identified this with the Jeans length.  It is possible that this length corresponds
to the width of a filament, but this is closely related to the Jeans length as
discussed above, and it is not clear how to distinguish these
possibilities from the MSDC.  In any event,
because the protostellar clouds are not generally rotationally
supported, collapse to a smaller structure will likely result in fragmentation
into binaries or multiple systems with separations
smaller than the original cloud Jeans length.  This provides an explanation of the
two breaks in the MSDC, one at the scale $\log \theta \sim -1.7$ or $r \sim$~0.05~pc
marking the appearance of binary systems, and the other at
0.25 pc corresponding to the Jeans length.\footnote{
It is worth noting that Figure 1b of
Simon (1997) suggests a similar break in the MSDC for Ophiuchus,
but at a slightly smaller scale than in Taurus.  This might indicate a smaller
Jeans length in this higher-density region (but see Bate, Clarke, \& McCaughrean
1998 for a discussion of some limitations of MSDC analyses).}
Implicit in this discussion is the idea that the stars have not dispersed far
from their birthsites, as suggested by the narrow distribution of Group 3.  
 
One problem in comparing the observations with a static model
is that the average velocity dispersion seen in the C$^{18}$O
cores ($\sim 0.4 \kms$ for starless regions) is considerably larger than
the sound speed at 10~K, $\sim 0.19 \kms$ (although in the densest regions
the velocity dispersion in the gas approaches the thermal value much more
closely; Rydbeck \etal 1977; Myers, Linke, \& Benson 1983;
Barranco \& Goodman 1999).  Thus, the equilibrium cylinder may
be used only as a guide to the true physical situation.  
However, the simulation of Nakajima \& Hanawa (1996) of filament formation
from a static sheet suggests that a relatively quiescent and static
filament might be present for some reasonable period of time;
in their model, supersonically-moving material which passes through a shock 
before accreting into the (relatively static) filament.

Onishi \etal (1998), who also proposed
that gravitational fragmentation is operating in the C$^{18}$O
filaments, suggested that the line
width should be interpreted in terms of an additional turbulent pressure
support.  If this were the case, changing the effective sound speed upward
by a factor of two would increase the characteristic length scale to a value
of order 1 pc, and the characteristic mass would be closer to $30 \msun$.
The situation might be even more complicated if a non-standard equation
of state were used in which turbulent pressure increases with decreasing
density (e.g., McLaughlin \& Pudritz 1997).  However, the observed velocity 
width may not represent simply pressure {\em support} but
more complex motions (Ballesteros-Paredes, Vazquez-Semadeni, \& Scalo 1999).
If the average line density in the filaments really does exceed the
critical value, the filaments could be collapsing radially, which might
explain part of the observed velocity dispersion.  Alternatively, one might
be detecting the supersonic motions of material accreting onto (relatively
static) filaments, as in the calculations of Nakajima \& Hanawa (1996).

Another problem is the efficiency of star formation.
It appears that only a fraction $\sim$~25\% of dense filament gas 
will become stars.  L\'eorat, Passot, \& Poquet (2000) and 
Klessen, Heitsch, \& Mac Low (2000) showed that small scale driving 
is needed to prevent gravitational collapse.  
Outflows or other protostellar energy input on small scales from
local sources, i.e. objects located within the filament, may be required
to disperse filamentary gas and limit star formation
(e.g., Adams \& Fatuzzo 1996; Matzner \& McKee 1999, 2000).

\subsection{Protostellar core structure}

This picture of gravitational fragmentation suggests that the fragmenting
cores might be elongated (e.g., Schneider \& Elmegreen 1979). 
The static model suggests aspect ratios of 2-3, as discussed above,
comparable to the average aspect ratio for optical cores of
2.4 found by Lee \& Myers (1999).  
Moreover, the cores show a tendency to be elongated along the filaments
(e.g., Myers \etal 1991), as would be predicted by the static model.
Figure 7 shows the spatial distribution and position
angles of all elongated optical cores studied by Lee \& Myers (1999).
The cores tend to be elongated along the filamentary structure
(Myers \etal 1991).  The alignment is particularly strong 
in the group 3/B18/L1506 region; only 2 of the 13 cores have orientations
deviating by more than about 30 degrees from the overall orientation
of the filament.  Note that Myers \etal (1991) showed that
many tracers - NH$_3$, extinction, etc. - show similar elongations on differing
scales, and so the choice of optical core properties does not bias the results.
It is also worth noting that Tachihara \etal (2000) found a similar trend
for cores to be aligned with filaments in Ophiuchus.

Again, one must emphasize that static models are not necessarily
applicable to the dynamic conditions in Taurus.  However,
the numerical simulations of Klessen \& Burkert (2000) found
that the gravitationally-unstable clumps were often elongated as part of
a larger filamentary structure, suggesting that the qualitative results
of the static stability analysis are relevant.

Some cores are aligned at large angles to the overall filamentary structure.
However, in several cases these objects
lie near more complex filamentary structure seen in $^{13}$CO 
(for example, the two cores near l=172, b=-15.5 are situated along a filamentary
section which runs nearly perpendicular to the overall filamentary structure,
as shown in Figure 3).  The more complex structure seen in the 
$^{13}$CO map suggests that small-scale structure
can play a non-negligible role in driving star formation.
Along these lines, it is worth noting that the filamentary structure near
the L1495 ``double group'' is quite complicated (cf. Figure 3), leading
to the speculation that intersecting smaller-scale flows have led to
building up a larger mass concentration which can then fragment into
groups that are denser than the other filaments (cf. Table 2), a feature
observed in the simulations of Klessen \& Burkert (2000, 2001) and Klessen
(2001).

Models of collapsing clouds indicate that close binary or multiple star formation
is best achieved by clouds which are not highly centrally concentrated initially,
and upon which some non-axisymmetric perturbation is imposed (Boss 1995,
and references therein; Boss 1997; Bodenheimer \etal 2000).  
In this regard the collapse of an elongated
cylinder might be nearly ideal for producing binaries (e.g., Boss 1993), 
which Taurus seems to do at a rate exceeding that of the field (Simon \etal 1995,
and references therein).
 
In a recent paper Jones \& Basu (2002; see also Jones, Basu, 
\& Dubinski 2001) suggest that cores are in general triaxial,
closer to prolate on large scales but more nearly oblate on small 
scales.\footnote{The opposite conclusion was erroneously reported in
the discussion in HBB.}  
It is difficult to reconcile the picture of oblate cores with the 
filamentary organization of Taurus cores and their clear systematic 
orientation along the filaments, especially in group 3.   
Fragmentation out of a filament could produce
oblate objects if contraction occurs along the filament; but then the
long axes of the cores would be aligned perpendicular to the filament,
not parallel as observed.  In this context, Curry (2002) has performed
an additional analysis of cores shapes and concluded that most are 
quasi-prolate, assuming random orientation of inclinations,
and has also pointed out the likelihood of fragmenting cores out of
filaments.

Alternatively, one might suppose that the
Taurus filaments are more elongated along the line of sight than their
minor axes projected upon the sky would indicate; in this way fragmentation
into long pieces along the filament could be consistent with       
quasi-oblate objects.  However, this would require that the sheets be
viewed nearly edge-on to appear filamentary. 
A test of this idea would be to study core orientations relative
to the larger-scale filamentary structure in other regions.  If cores are
systematically oriented closely along their natal filaments in most or
all regions, it is unlikely that this can be attributed to projection
effects, and would favor the quasi-prolate configuration.

It is also important to consider the evolution of a collapsing
flattened structure.  Consider for example the calculations
performed by Hartmann \etal (1994) for the isothermal self-gravitating
sheet.  In the early stages of gravitational collapse, 
a central condensation initially forms 
which is much rounder than the initial configuration - this is a natural product
of gravity operating on a scale of order the sheet scale height
or smaller.  This structure persists until material in the shortest dimension
has been able to fall in.  In other words, the cloud structure is not that of a uniform
(quasi-) ellipsoid on all scales.  This may be consistent with the
finding of Jones \& Basu (2002), who suggest that protostellar clouds
are more prolate on large scales even if more oblate on small scales.
A strong central density concentration which is
more nearly spherical does not mean that the parent core was not
highly elongated, nor that subsequent infall will be spherical.
Observations with higher spatial resolution, spanning a wider range of
densities, would be helpful in sorting among these possibilities.

\subsection{Dynamic star formation}

If protostellar cloud cores are generally (roughly) prolate,
at least on larger scales, this poses problems for
hydrostatic equilibrium models of cores supported primarily by magnetic
fields.  Curry \& Stahler (2001) have developed models of quasi-static cores which
are prolate; however, in these models the magnetic
field is oriented along the major axis of the cloud.  As noted in the
previous section, the large-scale magnetic field in Taurus is more
nearly perpendicular to the filaments and thus to the core elongations.
While there are no direct measurements of field directions in the filaments,
it is more plausible to suppose that the field remains roughly
perpendicular to the core major axes than otherwise.  
The model of Fiege \& Pudritz (2000), in which the filaments are constrained by
fields with a modest helical component, cannot be ruled out in the absence
of more detailed magnetic field measurements, but the maintainence of such
helical components seems to require special conditions, e.g., fixing the ends of the
helical magnetic field lines to maintain the twist.  
Curry (2000) has identified prolate equilibria in cylindrical geometry without
magnetic fields; however, it is not clear that such solutions are relevant,
given the observed high line densities and supersonic turbulence
observed in the Taurus filaments.

It seems more likely that the cores are not in hydrostatic equilibrium (Fleck 1992),
which would not require special conditions, though some regions may be relatively
quiescent or quasi-static.  Because the core lifetimes, of order a few
$\times 10^5$ yr (Lee \& Myers 1999), are consistent with the linear growth
times for gravitational collapse (\S 3.2), there is no need for exact hydrostatic
equilibrium.  It should be emphasized that if cores are gravitationally fragmenting
from (low velocity-dispersion) filaments, the very act of their formation implies
contraction on their largest scales, with subsonic motions during the linear growth
regime, consistent with the large-scale, slow infall inferred by Tafalla \etal (1998)
and Lee, Myers, \& Tafalla (2001).

In a dynamic picture, all that is required is
that shock dissipation of turbulent motions produces subsonic flows, 
in which cores can gravitationally fragment
(Padoan \& Nordlund 1999; Klessen \& Burkert 2000, 2001)).
Although cores are unlikely to be in hydrostatic equilibrium,
they may start their existence as subsonically-evolving objects in the
post-shock gas; for this reason static models of cores still can be used 
a guide to the more complex physics of real cores.

These considerations suggest the following, highly schematic,
picture of low-mass star formation
in Taurus (Figure 8).  Large-scale flows might form either filaments
or more likely sheet-like structures (Figure 8a), which becomes 
molecular when the accumulated surface density is large enough 
that $A_V \gtrsim 1$; at this point, self-gravity of the
layer becomes important (HBB).  An initially sheet-like configuration
might collapse laterally to form filaments, either due to turbulent
instabilities or gravitational contraction,
(Figure 8b, middle figure, side view).   
The turbulence driven by the flows has a large-scale component,
possibly associated with and/or channeled by the large-scale
magnetic field; this results in contraction into extended bands
or low density filaments.  The densest regions of these contracting
filaments, where secondary shocks have dissipated more of the
turbulent motions, gravitationally fragment into elongated structures 
which correspond to protostellar cloud cores (Figure 8b, bottom figure, top view).

This model is clearly highly schematic; Figure 3 shows that individual
dense filaments can be much more complex in structure than outlined
here.  Furthermore, even at a schematic level this might
not necessarily apply to all star formation in Taurus;
roughly 60\% of the total population do not fall into groups 1-4,
and the geometry of the more ``clustered'' groups (L1495, L1551)
is different.  On the other hand, it must be recognized that
dispersal of gas and stars, which has clearly gone on
in Taurus to some extent, will tend to erase any initial structure.
Even if the gaseous structure is more
complicated in general than long straight filaments (e.g., Figure 3),
the qualitative aspects of sheet formation by flows, flow-driven
or gravitational contraction into filaments due to anisotropic turbulent 
and magnetic pressures, followed by protostellar core fragmentation from filaments
may still be applicable.

\section{Conclusions}

The spatial distribution of young stars in Taurus is remarkably filamentary
and well-organized.  The large-scale coherence of the spatial distribution suggests 
that small-scale turbulence is not dominating the cloud structure,
consistent with driving by external flows which could have formed the cloud.  
The spatial distribution of young stars is consistent with a roughly 
uniform distribution along bands and filaments, with a spacing comparable 
to the local Jeans length.
The efficiency of star formation in filaments is much higher
than elsewhere; the higher frequency of protostars and accreting T Tauri stars 
indicates that the population in these filaments is relatively young.
The protostellar cores from which
the stars formed are often elongated along the filaments in which
they reside, consistent with formation by gravitational fragmentation.
Fragmentation of filaments naturally
produces elongated, roughly prolate cloud cores whose structure is well-suited
to producing binaries after collapse.  Numerical simulations of 
turbulent clouds with higher spatial resolution, and turbulent
driving on large scales, are urgently needed to compare with the observations
of this nearby region which constitutes the main example of
``quiescent'' star formation.  

The fragmentation of filaments into stars in Taurus has an echo on 
larger scales in regions of high mass and clustered star formation, 
a topic addressed in the second paper in this series.

The discussion of this paper was strongly influenced by my collaboration
with Javier Ballesteros, especially the many discussions we have had concerning
turbulent support.  The paper also benefited greatly from the suggestions,
comments, and encouragement of Phil Myers, who also suggested the importance
of the roughly periodic banded/filamentary structure of Taurus, and its implications
in terms of fragmentation from a sheet-like structure.  
This work was supported in part by NASA grant NAG5-9670.

\begin{figure}
\plotone{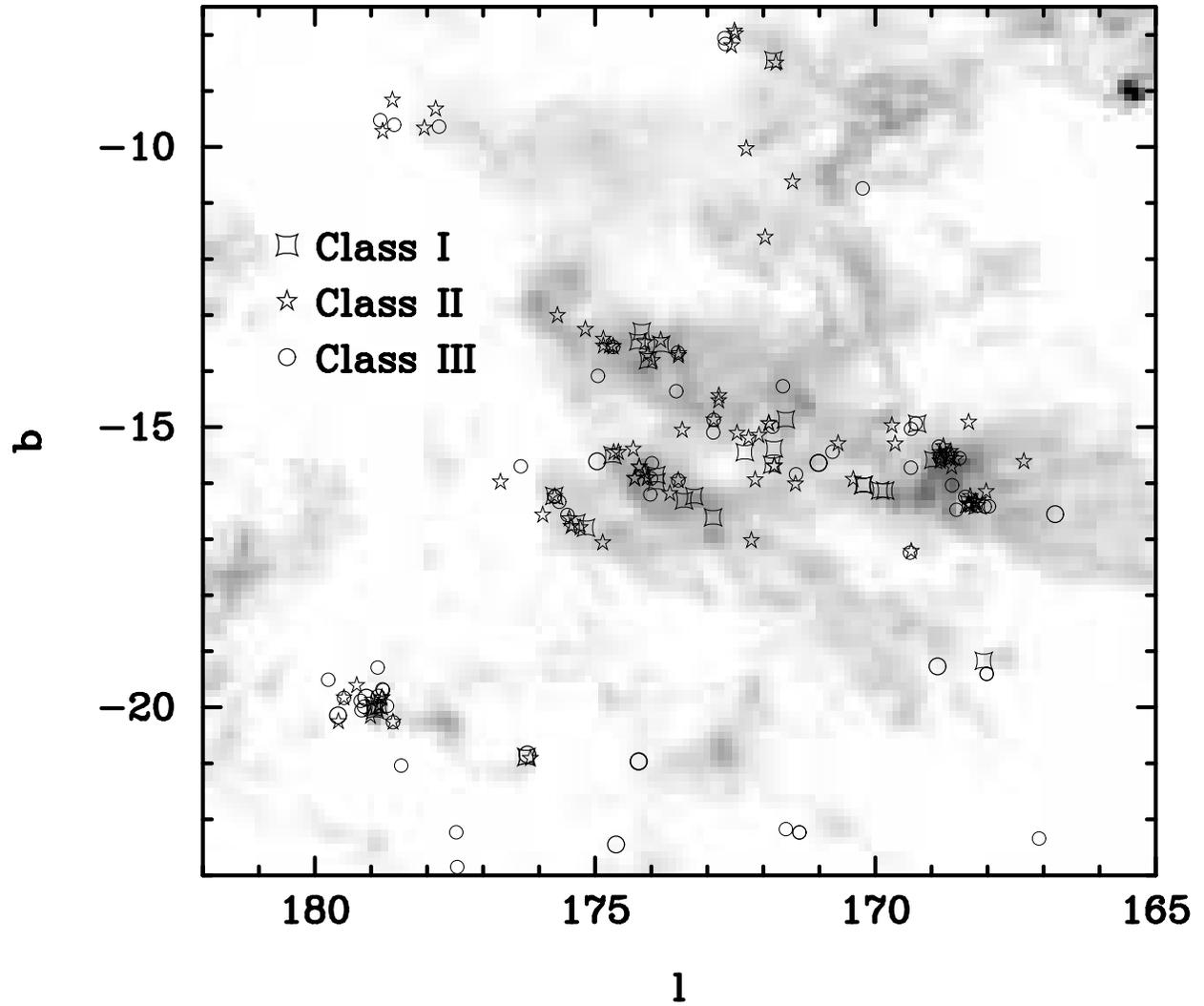}
\caption{Young stellar objects in the Taurus region, labelled by
their spectral energy distribution class, superimposed
upon the $^{12}$CO map of Megeath, Dame, \& Thaddeus (2002) (see text)}
\end{figure}

\begin{figure}
\plottwo{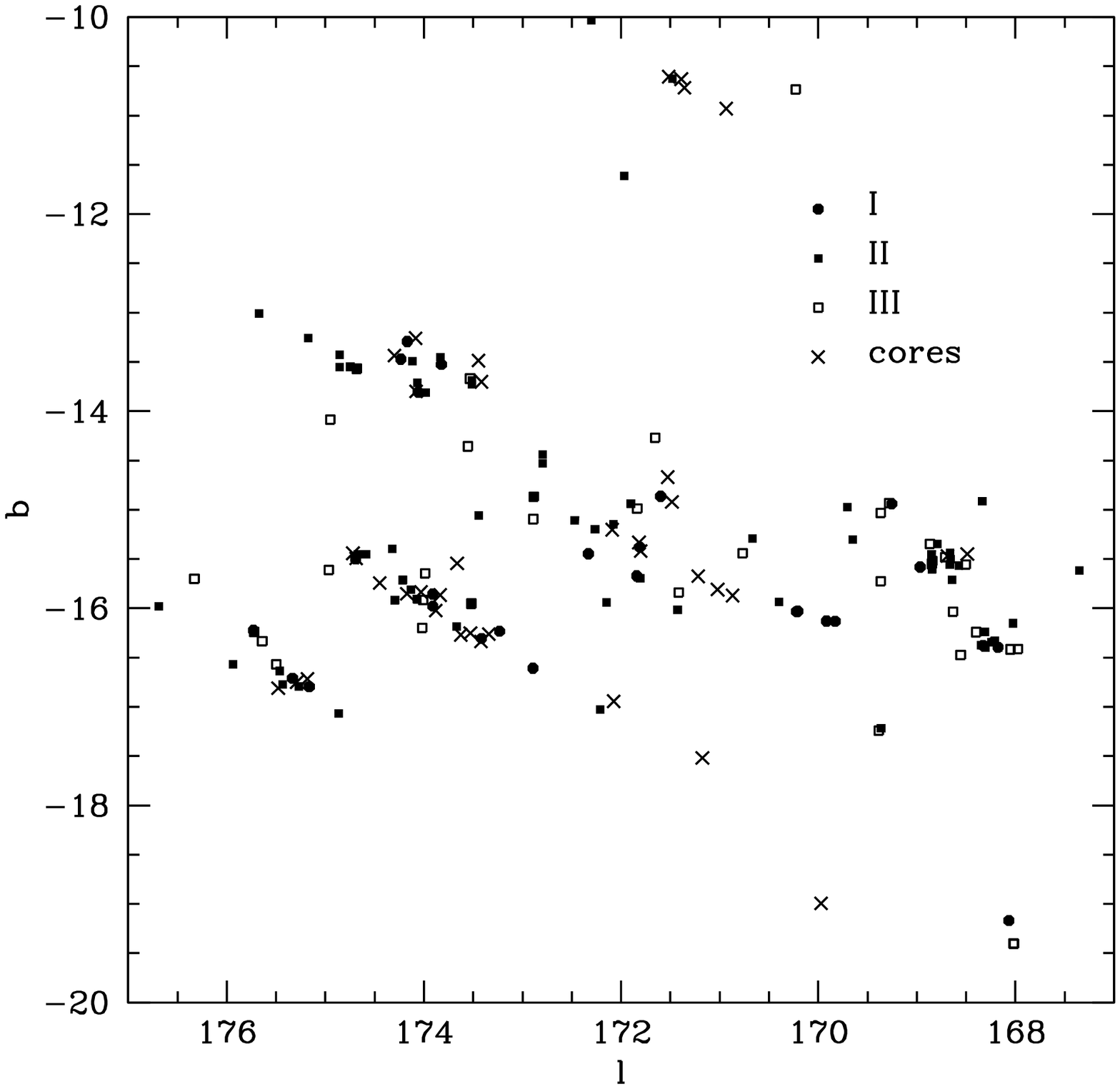}
{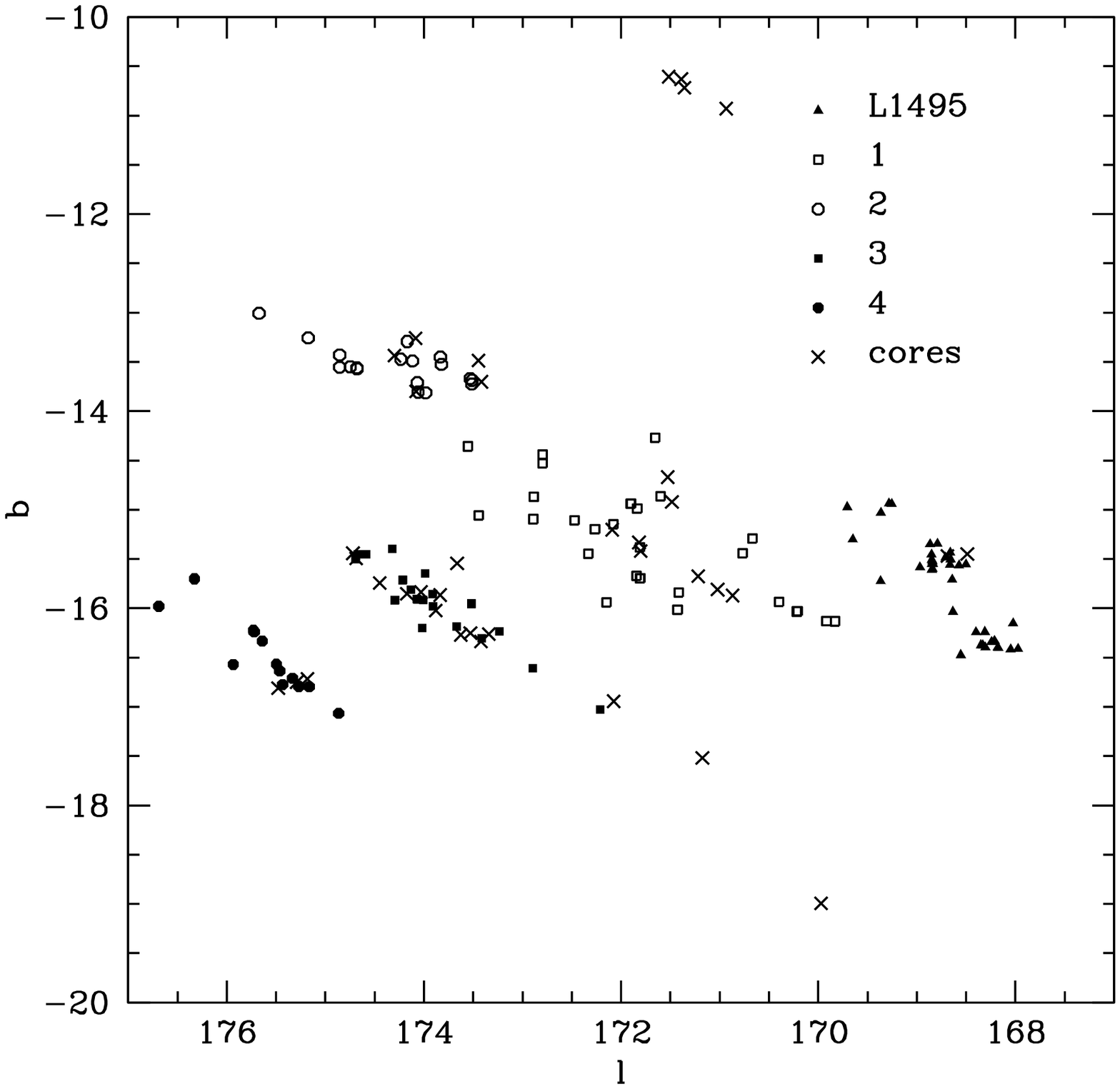}
\caption{ a) Spatial distribution of young stars, labelled
by their spectral energy distribution class, and optically-selected cores
in the central region of Figure 1;
b) Subset of stars in a) associated with groups, 
as labeled in the legend (see text; also Table 1)}
\end{figure}

\begin{figure}
\plotone{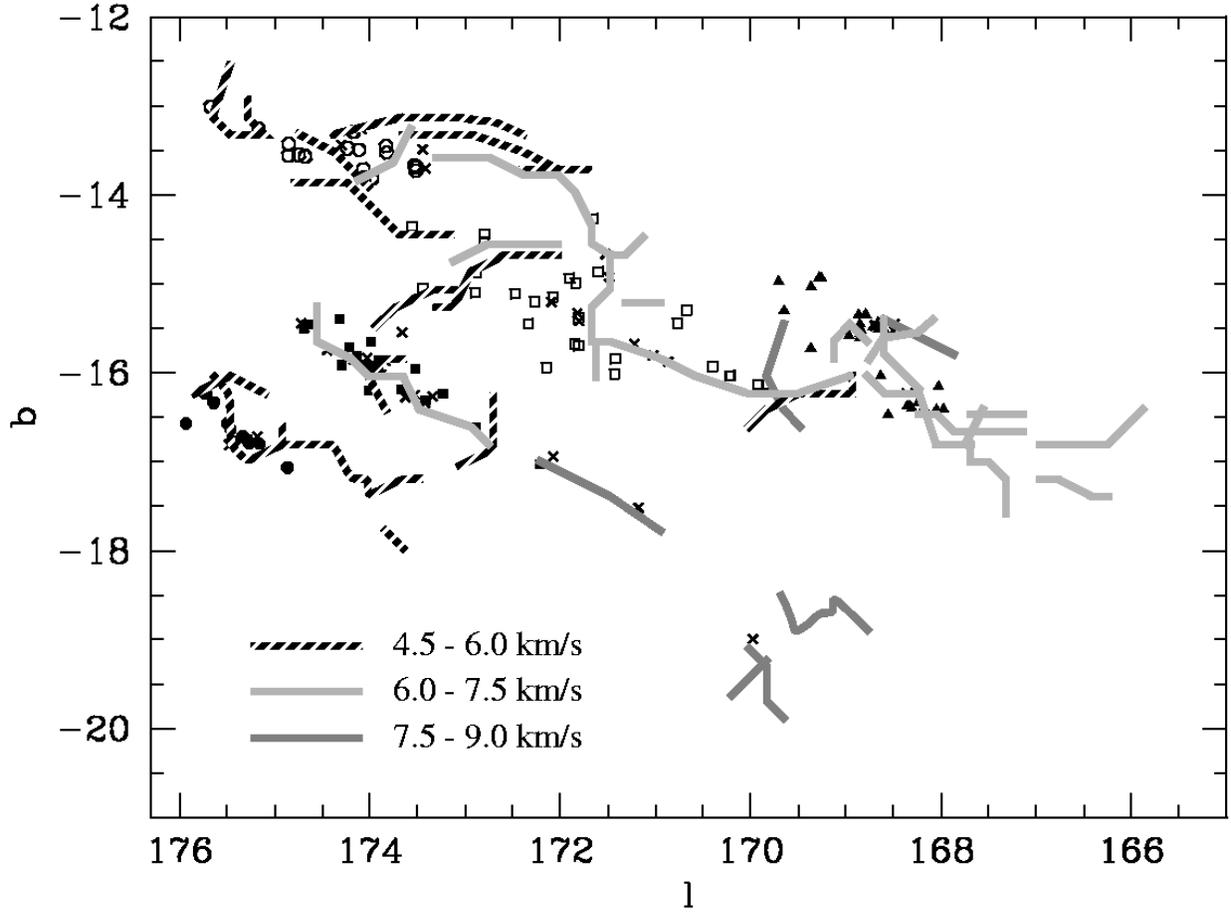}
\caption{Approximate position of $^{13}$CO filaments in
the central region of Taurus found by Mizuno \etal (1995),
superimposed upon the groups of Figure 2b.  The CO filaments
are sorted into the same LSR velocity range as shown by Mizuno \etal.}
\end{figure}

\begin{figure}
\plotone{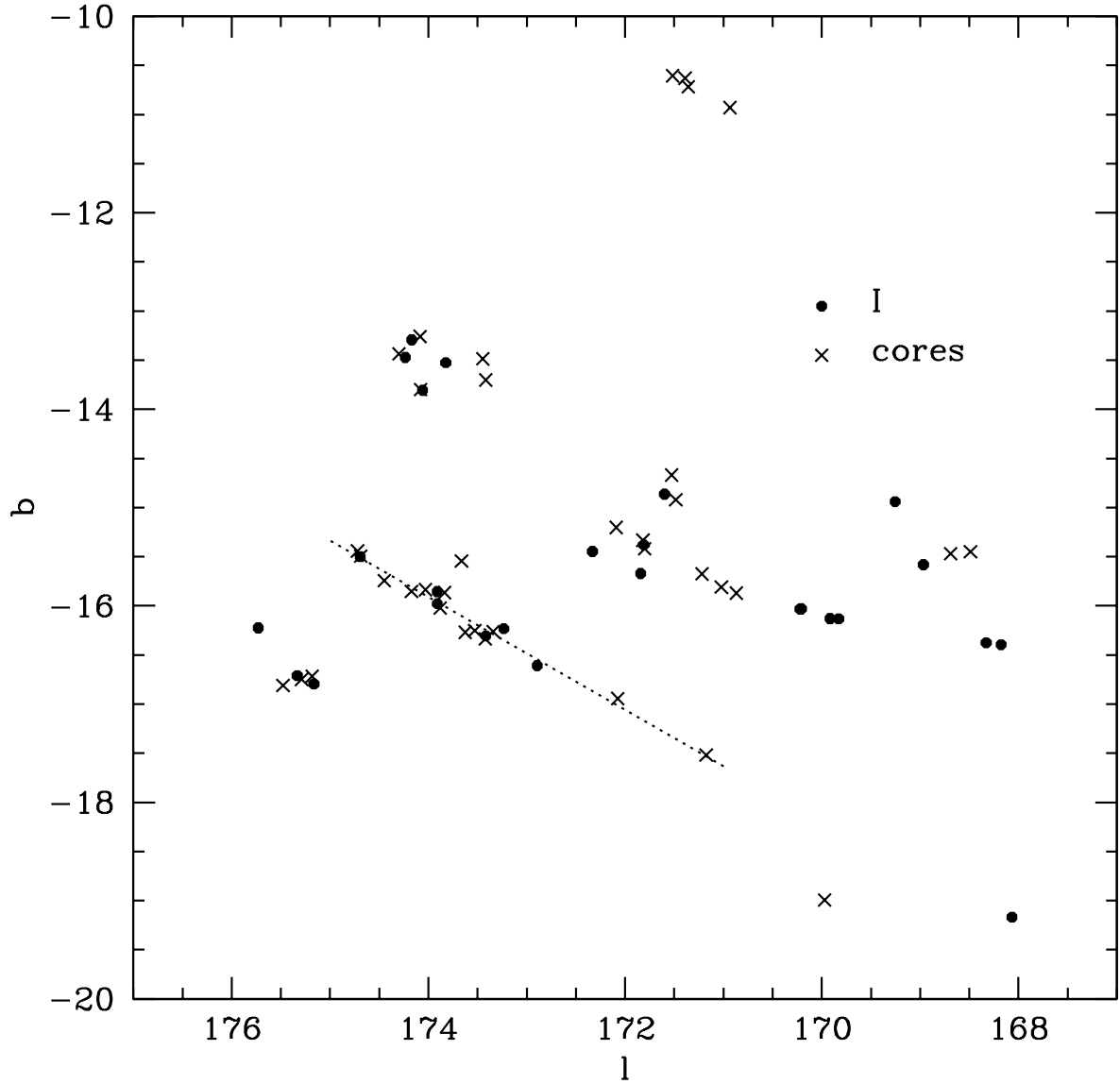}
\caption{Spatial distribution of Class I objects and optical cores from
the survey of Lee \& Myers (1999).  The dotted line is a least squares
linear fit to the objects in the region of group 3 (see text)}
\end{figure}

\begin{figure}
\plotone{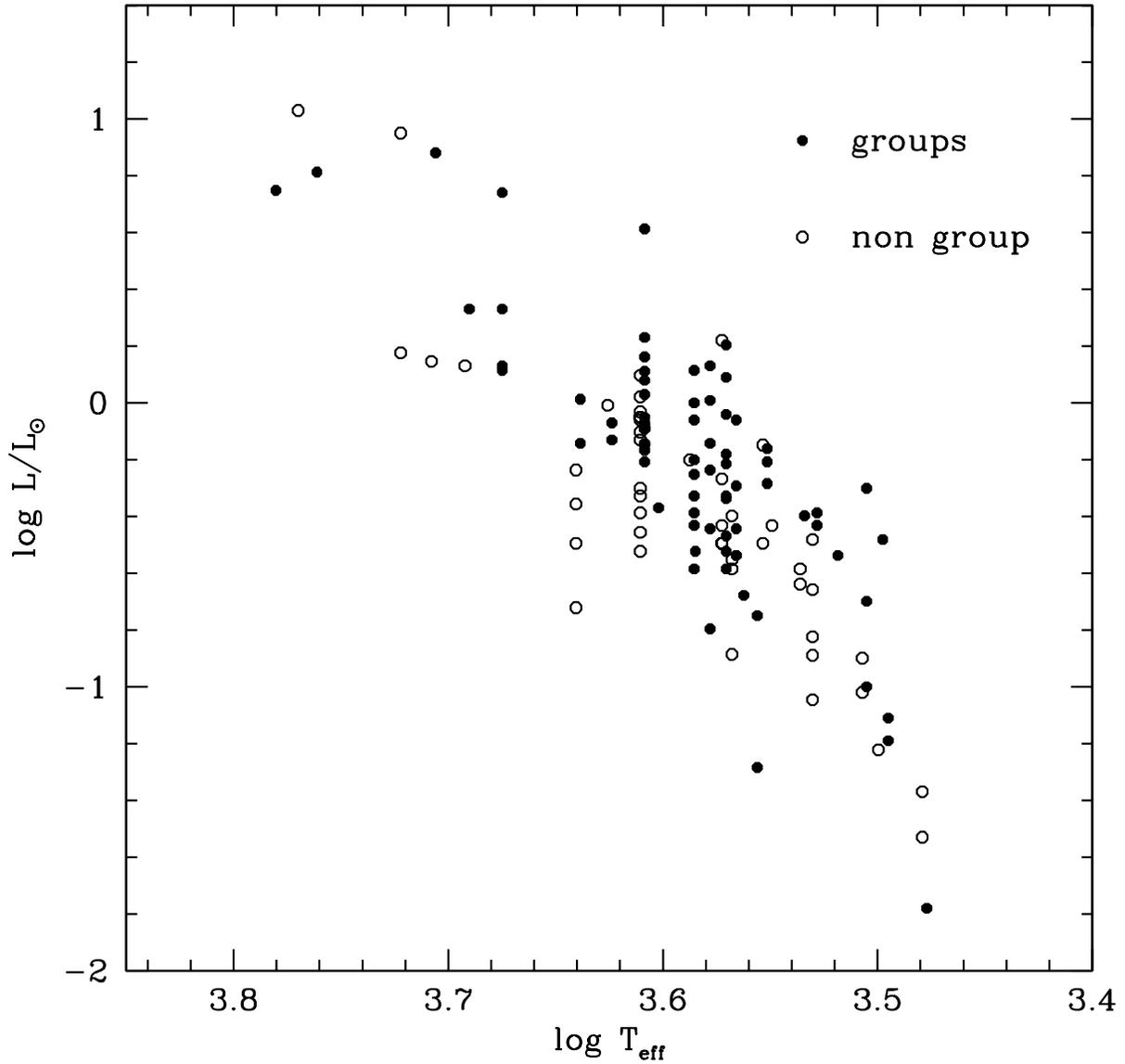}
\caption{HR diagram for stars with available stellar luminosities
and effective temperatures (see text), sorted between members of
the groups and others.  The group stars appear to be somewhat
younger than the non-group objects, though the result is marginal
due to possible systematic errors (see text) }
\end{figure}

\begin{figure}
\plotone{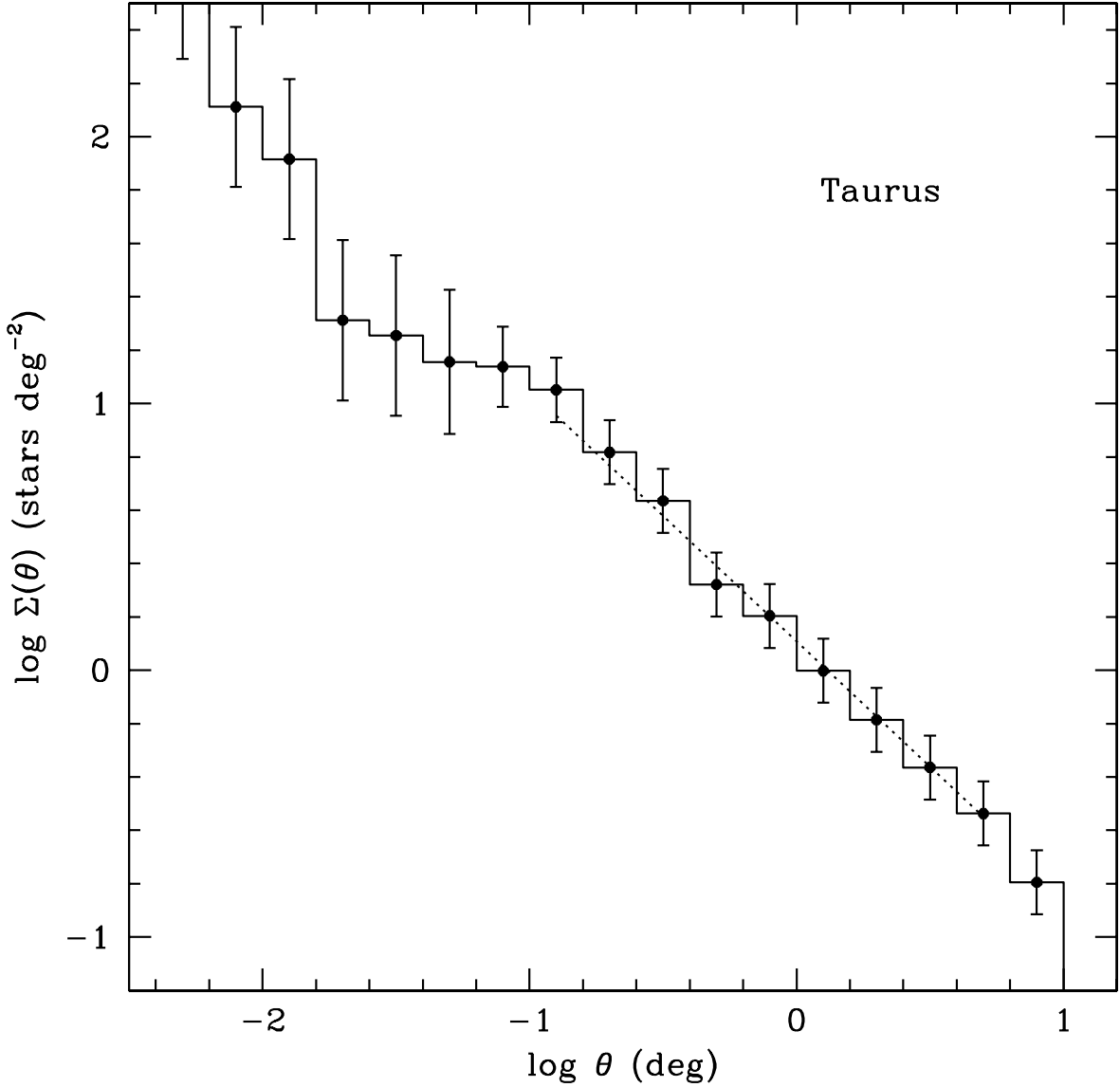}
\caption{The mean surface density of companions (MSDC), $\Sigma(\theta)$, 
as a function of angular separation $\theta$ (degrees).
The dotted line is a least-squares linear
fit with slope $-0.98$ to the bins between $-0.9 \leq \log \theta \leq 0.7$.
At small separations the MSDC
rises rapidly due to binary and multiple companions 
The break in slope at $\log \theta \sim -1$
corresponds to a spatial separation of about 0.24 pc at the distance
of Taurus.}
\end{figure}

\begin{figure}
\plotone{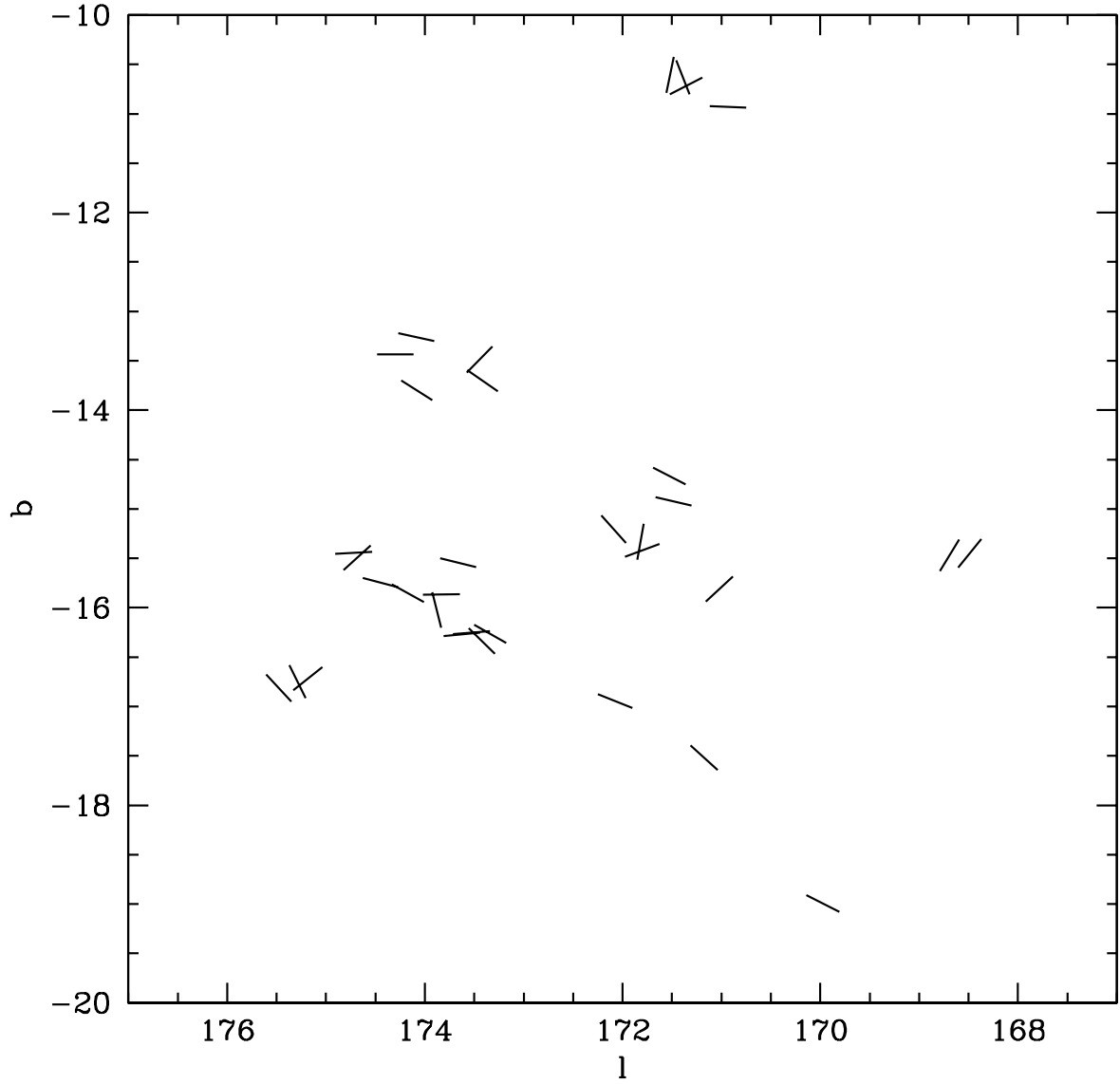}
\caption{Spatial distribution of optical cores from Lee \& Myers (1999),
with the orientation of the line indicating the position angle of
the core major axis.}
\end{figure}

\begin{figure}
\plotone{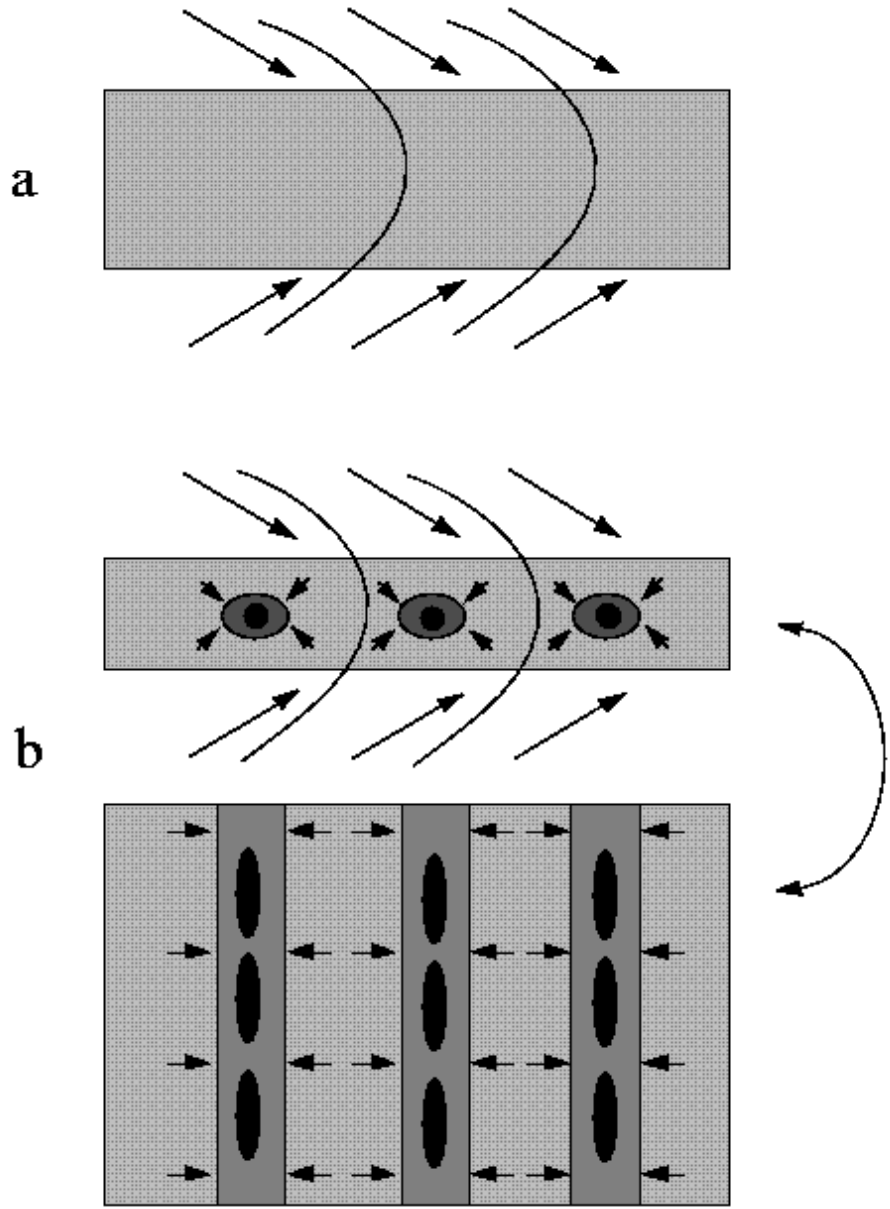}
\caption{Suggested outline of star formation in Taurus filaments. Converging
large-scale flows in the interstellar medium form a sheet-like structure
(a).  Magnetic fields may help channel the flow.  When sufficient column
density is amassed, sheet is subject to gravitational fragmentation into
filaments (b) (e.g., Schneider \& Elmegreen 1979; Miyama \etal 1987a,b; 
Nakajima \& Hanawa 1996), possibly aided by
anisotropic (large-scale) turbulence and/or magnetic field channelling.
The collapsing filaments develop dense cores (top view, bottom part of (b))
which then fragment into protostellar cores (see text).}
\end{figure}

\newpage

\begin{deluxetable}{rrrrrr} 
\tablecolumns{6} 
\tablewidth{0pc} 
\tablecaption{Taurus Groups} 
\tablehead{ 
\colhead{l} & \colhead{b}   & \colhead{RA(1950)}    & \colhead{Dec(1950)} & 
\colhead{Name}    & \colhead{Class}}
\startdata 
\cutinhead{Group 1}
169.8285 & -16.1320 & 04:16:38.30 & 27:06:23.0 & 04166+2706 & I \\
169.9167 & -16.1288 & 04:16:54.60 & 27:02:49.0 & 04169+2702 & I \\
170.2079 & -16.0303 & 04:18:05.50 & 26:54:37.0 & 04181+2655 & I \\
170.2185 & -16.0327 & 04:18:06.90 & 26:54:04.0 & 04181+2654A & I \\
170.2185 & -16.0327 & 04:18:06.90 & 26:54:04.0 & 04181+2654B & I \\
171.8103 & -15.3798 & 04:24:53.70 & 26:12:41.0 & 04248+2612 & I \\
172.3320 & -15.4465 & 04:26:09.80 & 25:47:30.0 & 04361+2547 & I \\
171.5982 & -14.8631 & 04:26:00.30 & 26:42:31.0 & 04260+2642 & I \\
170.3973 & -15.9353 & 04:18:57.60 & 26:50:30.5 & FS Tau & II \\
170.6678 & -15.2938 & 04:21:52.10 & 27:05:08.0 & IP Tau & II \\
171.4271 & -16.0164 & 04:21:41.30 & 26:03:30.0 & 04216+2603 & II \\
171.8052 & -15.6962 & 04:23:49.90 & 26:00:13.0 & FV Tau & II \\
171.8080 & -15.6946 & 04:23:50.70 & 26:00:09.7 & FV Tau/c & II \\
171.8424 & -15.6720 & 04:24:01.10 & 25:59:35.5 & DG Tau & II \\
171.8990 & -14.9386 & 04:26:37.10 & 26:26:28.9 & DH Tau & II \\
171.9033 & -14.9377 & 04:26:38.00 & 26:26:20.0 & DI Tau & II \\
172.0771 & -15.1492 & 04:26:25.55 & 26:10:22.6 & FW Tau=CIDA4 & II \\
172.1465 & -15.9423 & 04:23:59.60 & 25:35:41.7 & DF Tau & II \\
172.2647 & -15.1990 & 04:26:47.70 & 26:00:16.0 & IQ Tau & II \\
172.4695 & -15.1082 & 04:27:40.80 & 25:54:59.0 & DK Tau & II \\
172.7942 & -14.4392 & 04:30:50.50 & 26:07:14.0 & IT Tau=J 644 & II \\
172.7966 & -14.5294 & 04:30:32.70 & 26:03:34.5 & IS Tau & II \\
172.8864 & -14.8703 & 04:29:39.30 & 25:46:13.0 & UZ Tau e & II \\
173.4457 & -15.0568 & 04:30:36.00 & 25:14:24.0 & DL Tau & II \\
171.8363 & -14.9900 & 04:26:16.00 & 26:27:09.0 & J 507 & III \\
170.7675 & -15.4414 & 04:21:40.30 & 26:54:54.0 & J 4423 & III \\
171.4156 & -15.8420 & 04:22:13.80 & 26:11:02.0 & J 4872 & III \\
171.6520 & -14.2686 & 04:28:09.10 & 27:03:54.0 & JH 56 & III \\
172.8864 & -14.8703 & 04:29:39.30 & 25:46:13.0 & UZ Tau w & III \\
172.8913 & -15.0952 & 04:28:55.00 & 25:37:08.0 & J 665 & III \\
173.5555 & -14.3572 & 04:33:15.40 & 25:36:55.0 & LkCa 14 & III \\
\cutinhead{Group 2}
173.8235 & -13.5257 & 04:36:49.30 & 25:57:16.0 & 04368+2557 & I \\
174.0586 & -13.8065 & 04:36:31.20 & 25:35:56.0 & 04365+2535 & I \\
174.2349 & -13.4732 & 04:38:08.50 & 25:40:53.0 & 04381+2540 & I \\
173.5117 & -13.6870 & 04:35:24.20 & 26:04:55.0 & DO Tau & II \\
173.5146 & -13.7251 & 04:35:16.91 & 26:03:18.6 & GM Tau=CIDA5 & II \\
173.8322 & -13.4512 & 04:37:06.00 & 25:59:45.0 & 04370+2559 & II \\
173.9804 & -13.8124 & 04:36:16.98 & 25:39:11.5 & GN Tau & II \\
174.0667 & -13.7120 & 04:36:51.80 & 25:39:13.0 & IC 2087 IR & II \\
174.1151 & -13.4909 & 04:37:45.00 & 25:45:34.0 & JH 223 & II \\
174.1697 & -13.2929 & 04:38:34.60 & 25:50:44.0 & 04385+2550 & II \\
174.6803 & -13.5606 & 04:39:04.20 & 25:17:32.8 & V955 Tau & II \\
174.6803 & -13.5645 & 04:39:03.40 & 25:17:23.7 & CoKu LkH332G1 & II \\
174.7475 & -13.5498 & 04:39:17.45 & 25:14:56.1 & CIDA-7 & II \\
174.8543 & -13.4296 & 04:39:59.60 & 25:14:43.0 & GO Tau & II \\
174.8544 & -13.5532 & 04:39:34.30 & 25:09:59.9 & DP Tau & II \\
175.1741 & -13.2561 & 04:41:27.50 & 25:06:53.0 & 04414+2506 & II \\
175.6718 & -13.0074 & 04:43:39.54 & 24:53:42.60 & RXJ04467+2459 & II \\
173.5306 & -13.6696 & 04:35:30.90 & 26:04:45.2 & HV Tau & III \\
174.6784 & -13.5698 & 04:39:02.00 & 25:17:16.8 & CoKu LkH332/G2 & III \\
\cutinhead{Group 3}
172.8944 & -16.6094 & 04:23:54.50 & 24:36:54.0 & 04239+2436 & I \\
173.2329 & -16.2331 & 04:26:05.70 & 24:37:17.0 & 04264+2433 & I \\
173.4157 & -16.3056 & 04:26:21.90 & 24:26:30.0 & GV Tau & I \\
174.6927 & -15.5004 & 04:32:33.50 & 24:02:15.0 & 04325+2402 & I \\
173.9105 & -15.8558 & 04:29:13.60 & 24:22:42.9 & Haro 6-13 & I/II \\
173.9094 & -15.9788 & 04:28:48.90 & 24:17:56.0 & HK Tau & I/II \\
172.2120 & -17.0264 & 04:20:37.20 & 24:49:20.0 & FT Tau & II \\
173.5194 & -15.9525 & 04:27:49.30 & 24:35:56.9 & ZZ Tau IRS & II \\
173.6685 & -16.1851 & 04:27:27.90 & 24:20:18.0 & FX Tau & II \\
174.0690 & -15.9106 & 04:29:28.90 & 24:13:38.6 & FY Tau & II \\
174.0709 & -15.9062 & 04:29:30.10 & 24:13:44.0 & FZ Tau & II \\
174.1313 & -15.8091 & 04:29:59.43 & 24:14:53.54 & MHO-8 & II \\
174.2137 & -15.7127 & 04:30:32.30 & 24:15:03.0 & GI Tau & II \\
174.2172 & -15.7135 & 04:30:32.70 & 24:14:52.0 & GK Tau & II \\
174.2939 & -15.9146 & 04:30:05.20 & 24:03:39.0 & V807 Tau & II \\
174.2974 & -15.9196 & 04:30:04.80 & 24:03:18.0 & GH Tau & II \\
174.3207 & -15.3951 & 04:31:53.50 & 24:22:44.0 & AA Tau & II \\
174.5856 & -15.4517 & 04:32:25.70 & 24:08:52.0 & DN Tau & II \\
174.6716 & -15.4528 & 04:32:39.60 & 24:05:01.8 & CoKu Tau/3 & II \\
173.5194 & -15.9525 & 04:27:49.30 & 24:35:56.9 & ZZ Tau & III \\
173.9893 & -15.6471 & 04:30:08.30 & 24:27:26.7 & V830 Tau & III \\
174.0062 & -15.9172 & 04:29:17.20 & 24:16:07.5 & V928 Tau & III \\
174.0170 & -16.2016 & 04:28:22.40 & 24:04:29.7 & V927 Tau & III \\
\cutinhead{Group 4}
175.1648 & -16.7957 & 04:29:32.20 & 22:51:11.0 & 04295+2251 & I \\
175.3343 & -16.7109 & 04:30:16.40 & 22:47:04.0 & 04302+2247 & I \\
175.7293 & -16.2243 & 04:32:56.80 & 22:48:31.0 & Haro 6-28 & I \\
174.8657 & -17.0661 & 04:27:50.20 & 22:53:40.0 & 04278+2253 & II \\
175.2680 & -16.7936 & 04:29:49.30 & 22:46:45.0 & JH 112 & II \\
175.4321 & -16.7747 & 04:30:19.50 & 22:40:18.0 & 04303+2240 & II \\
175.4641 & -16.6360 & 04:30:52.20 & 22:44:16.7 & CI Tau & II \\
175.7222 & -16.2382 & 04:32:52.90 & 22:48:17.7 & HP Tau & II \\
175.9349 & -16.5703 & 04:32:20.80 & 22:26:06.6 & HO Tau & II \\
176.6895 & -15.9814 & 04:36:18.40 & 22:15:12.7 & LkCa 15 & II \\
175.4964 & -16.5670 & 04:31:11.10 & 22:45:32.0 & JH 108 & III \\
175.6401 & -16.3325 & 04:32:20.90 & 22:48:16.8 & FF Tau & III \\
175.7222 & -16.2382 & 04:32:52.90 & 22:48:17.7 & HP Tau/G2 & III \\
175.7222 & -16.2382 & 04:32:52.90 & 22:48:17.7 & HP Tau/G3 & III \\
176.3280 & -15.7005 & 04:36:17.40 & 22:42:02.0 & VY Tau & III \\
\cutinhead{L 1495}
168.1774 & -16.3955 & 04:10:49.30 & 28:03:57.0 & 04108+2803 B & I \\
168.0247 & -16.1523 & 04:11:08.60 & 28:20:26.9 & FN Tau & II \\
168.1756 & -16.4005 & 04:10:48.00 & 28:03:49.0 & 04108+2803 A & II \\
168.2101 & -16.3310 & 04:11:07.80 & 28:05:18.8 & FM Tau & II \\
168.2166 & -16.3397 & 04:11:07.30 & 28:04:41.0 & V773 Tau & II \\
168.2426 & -16.3436 & 04:11:11.30 & 28:03:27.0 & CW Tau & II \\
168.3044 & -16.3975 & 04:11:12.09 & 27:58:38.6 & CIDA-1 & II \\
168.3096 & -16.2403 & 04:11:43.60 & 28:05:01.6 & FO Tau & II \\
168.3288 & -16.3761 & 04:11:20.71 & 27:58:32.46 & MHO-1 & II \\
168.3289 & -16.3757 & 04:11:20.80 & 27:58:33.0 & 04113+2758 & II \\
168.3295 & -16.3762 & 04:11:20.81 & 27:58:30.46 & MHO-2 & II \\
168.3500 & -16.3743 & 04:11:24.92 & 27:57:44.73 & MHO-3 & II \\
168.5722 & -15.5660 & 04:14:43.35 & 28:22:18.5 & CIDA-3 & II \\
168.6409 & -15.7095 & 04:14:27.70 & 28:13:28.6 & CY Tau & II \\
168.6559 & -15.5033 & 04:15:10.93 & 28:21:25.53 & v410a13 & II \\
168.6626 & -15.4350 & 04:15:25.60 & 28:23:59.0 & 04154+2823 & II \\
168.8353 & -15.5492 & 04:15:34.50 & 28:12:01.8 & V892 Tau & II \\
168.8406 & -15.5993 & 04:15:25.60 & 28:09:44.0 & CZ Tau & II \\
168.7902 & -15.3457 & 04:16:06.40 & 28:22:21.0 & FQ Tau & II \\
168.8454 & -15.6063 & 04:15:25.10 & 28:09:14.6 & DD Tau & II \\
168.8490 & -15.5064 & 04:15:45.40 & 28:13:14.0 & CoKu Tau/1 & II \\
168.8499 & -15.4543 & 04:15:55.82 & 28:15:21.47 & v410xr5a & II \\
168.9670 & -15.5824 & 04:15:51.80 & 28:05:09.0 & 04158+2805 & II \\
169.2547 & -14.9396 & 04:18:50.90 & 28:19:35.0 & RY Tau & II \\
169.6484 & -15.3018 & 04:18:49.80 & 27:48:05.0 & DE Tau & II \\
169.7051 & -14.9734 & 04:20:05.10 & 27:59:08.0 & 04200+2759 & II \\
168.6649 & -15.5575 & 04:15:01.90 & 28:18:48.0 & v410xr3 & II \\ 
167.9731 & -16.4132 & 04:10:08.46 & 28:11:35.2 & LkCa 1 & III \\
168.0482 & -16.4167 & 04:10:21.54 & 28:08:21.5 & Anon 1 & III \\
168.3986 & -16.2417 & 04:11:59.57 & 28:01:17.8 & CIDA-2 & III \\
168.5014 & -15.5550 & 04:14:32.62 & 28:25:42.6 & LkCa 5 & III \\
168.5554 & -16.4745 & 04:11:42.80 & 27:45:05.0 & LkCa 3 & III \\
168.6330 & -16.0356 & 04:13:22.40 & 28:00:13.0 & LkCa 4 & III \\
168.7099 & -15.4826 & 04:15:24.83 & 28:20:01.7 & V410 Tau & III \\
168.8413 & -15.5222 & 04:15:40.90 & 28:12:54.0 & Hubble 4 & III \\
168.8460 & -15.5488 & 04:15:36.52 & 28:11:36.20 & v410xr7 & III \\
168.8670 & -15.3480 & 04:16:19.90 & 28:19:02.6 & V819 Tau & III \\
169.2815 & -14.9352 & 04:18:56.60 & 28:18:37.7 & LkCa 21 & III \\
169.3654 & -15.0320 & 04:18:52.50 & 28:11:06.6 & HD 283572 & III \\
169.3675 & -15.7256 & 04:16:35.80 & 27:42:28.0 & LkCa 7 & III \\
\enddata 
\end{deluxetable} 

\newpage

\begin{deluxetable}{lccrrr}
\tablecolumns{6}
\tablewidth{0pc} 
\tablecaption{Density of Taurus groups \label{tbl-2}}
\tablehead{ \colhead{Group} & \colhead{N(YSO)} & \colhead{N(main)\tablenotemark{a}} 
&  \colhead{l(deg)} & \colhead N/l(deg) & \colhead N/l(pc)}
\startdata
1     &31    &31          &4.01  &7.7  & 3.2 \\
2     &19    &18          &1.67  &10.8 & 4.4  \\
1$+$2 &50    &49          &5.92  &8.3  &3.4 \\
3     &23    &22          &2.06 &10.7  &4.4 \\
4     &15    &13          &1.17 &11.1  &4.5 \\
L1495 &40    &40          &2.20 &18.2  &7.4 \\
\enddata

\tablenotetext{a}{Total number of stars considered, with outlying
objects eliminated (see text):  easternmost object in Group 2,
westernmost object in Group 3, and the two easternmost members
of Group 4 were eliminated.}

\end{deluxetable}
\end{document}